\documentclass[twoside,slac_one]{revtex4}
\usepackage{a4p}
\usepackage{atlasphysics}
\usepackage{macross}

\usepackage{graphicx}
\usepackage{fancyhdr}
\usepackage{amsmath} 
\usepackage{bm}
\usepackage{amsxtra}
\usepackage{amssymb}
\usepackage{amsthm}
\usepackage{latexsym}
\usepackage{lscape}

\usepackage{booktabs}
\usepackage{xspace}

\begin{document}

\title{Measurement of the $W^+W^-$ Cross Section in $\sqrt{s}=7$ TeV pp Collisions with the ATLAS Detector}

%

\author{Haijun Yang (on behalf of the ATLAS Collaboration)}
\affiliation{Department of Physics, University of Michigan, Ann Arbor, MI 48109-1120, USA}

\begin{abstract}
We report a measurement of the \WW production cross section 
in $pp$ collisions at $\sqrt s = $7~\TeV.
The \WW leptonic decay channels are analyzed using 
data corresponding to 1.02 \ifb of integrated luminosity
collected by the ATLAS detector
during 2011 at the CERN Large Hadron Collider.
With 414 observed \WW\ candidate events and an estimated background of 170$\pm$28 events, 
the measured \WW production cross section is
$48.2 \pm 4.0 {\rm(stat)} \pm 6.4 {\rm(syst)} \pm 1.8 {\rm(lumi)}$ pb,
in agreement with the Standard Model NLO prediction of 46$\pm$3~pb.
\end{abstract}

\maketitle

\thispagestyle{fancy}


\section{Introduction}

\indent
The measurement of the $W^+W^-$ production cross section at the LHC provides an
important test of the Standard Model (SM) through the sensitivity to the 
triple gauge boson 
couplings that result from the non-Abelian structure of the gauge symmetry group, 
$SU(2)_{\mathrm{L}}\times U(1)_{\mathrm{Y}}$.  
Furthermore, non-resonant $W^+W^-$ production is an irreducible background 
in searches for the Higgs boson in the same final state.  
Understanding the detection sensitivity of ATLAS to $W$-pair production 
is crucial for Higgs boson searches.
The WW cross section has been measured at LHC with the ATLAS detector 
using limited statistics based on 2010 data (34 \ipb)~\cite{atlas-ww2010}, 
this paper presents new results with 1.02 \ifb integrated luminosity 
collected in 2011.



Candidate \WW events are reconstructed in the leptonic decay
channels, $W^\pm\rightarrow\ell^\pm \nu$, where $\ell$ is an electron or muon. 
Sequential decays to electrons and muons via $\tau$ leptons are also included as signal:
$W^\pm\rightarrow\tau^\pm\nu_\tau \rightarrow \ell^\pm\nu_\ell \nu_\tau\bar{ \nu_\tau}$. 
The resulting final state has two high-transverse-momentum (high-\pt) charged
leptons and substantial transverse momentum imbalance in the final state (referred to as 
missing transverse energy \met below) due to the neutrinos or antineutrinos escaping detection.
The major backgrounds for this $\ell^+\ell^- \met $ final state are 
Drell-Yan, top quark, $W+$jet and diboson ($WZ$, $ZZ$ and $W\gamma$) production.
%
%
\section{The ATLAS detector and data sample}

\indent
The ATLAS detector~\cite{detectorpaper} is a multipurpose particle
physics apparatus with nearly 4$\pi$ solid angle coverage.
ATLAS uses a right-handed coordinate system with its origin at the
nominal interaction point (IP) in the centre of the detector and the
$z$-axis along the beam pipe. The $x$-axis points from the IP to the
centre of the LHC ring, and the $y$ axis points upward. 
Cylindrical coordinates $(r,\phi)$ are used in the transverse plane, $\phi$ being
the azimuthal angle around the beam pipe. The pseudorapidity is
defined in terms of the polar angle $\theta$
as $\eta = -\ln [\tan (\theta/2)]$. Closest to the beamline are
inner tracking detectors which use layers of silicon-based and
straw-tube detectors, located inside a thin superconducting
solenoid that provides a 2T magnetic field, to measure
the trajectories of charged particles within $|\eta|<2.5$. The solenoid is
surrounded by a hermetic calorimeter system. A liquid argon
(LAr) sampling calorimeter is divided into a central
barrel calorimeter and two end-cap calorimeters, each
housed in a separate cryostat. Fine-grained LAr electromagnetic
(EM) calorimeters, with excellent energy resolution,
provide coverage for $|\eta| < $3.2. 
An iron-scintillator tile calorimeter provides hadronic coverage in the range
$|\eta| < $1.7. In the end-caps ($|\eta| > $1.5), LAr hadronic
calorimeters match the outer $|\eta|$ limits of the end-cap
EM calorimeters. LAr forward calorimeters provide both
EM and hadronic energy measurements, and extend the
coverage to $|\eta| < $4.9. Outside the calorimeters is an
extensive muon spectrometer in a toroidal magnetic field, 
providing precise muon measurements within $|\eta| <$ 2.7. 
The muon trigger system covers the range $|\eta| < $2.4.

A three-level trigger system is used to select interesting events in real time. 
The electron trigger selects electrons that deposit 
at least 20 GeV of transverse energy in the calorimeter.
The muon trigger requirement consists of a logical OR between a
trigger that requires a muon detected with a transverse momentum of at
least 18 GeV and a looser quality trigger that requires a muon of at
least 40 GeV in the barrel section of the muon spectrometer.
The data used for this analysis were recorded up to June 2011 and correspond to an integrated luminosity of 1.02 fb$^{-1}$. The luminosity 
is determined with a relative uncertainty of 3.7\% using
van der Meer scans~\cite{ATLAS-CONF-2011-011}.

%
%



%
%


\indent
Monte Carlo simulation samples are used to develop and validate the analysis procedures, 
to calculate the acceptance for \WW events and to evaluate the contributions 
from some background processes. 

For the \WW signal, we use the next-to-leading order (NLO) generator
\mcatnlo~\cite{MCatNLO} in conjunction with the CTEQ6.6~\cite{CTEQ66}
parton distribution functions (PDF).
\herwig~\cite{herwig} is used for $W$ leptonic decays and for the parton shower simulation and hadronization. 
\Jimmy~\cite{Jimmy} is used for the underlying event simulation. 
In addition, the \ggtwoww~\cite{Binoth:2006mf} MC generator (interfaced to \herwig and \Jimmy) 
is used to simulate the \WW events from the gluon-gluon fusion process. 

The Drell-Yan and the $W+$ jets backgrounds are generated with \alpgen.
The top ($t\bar t$ and $Wt$) and diboson ($WZ$ and $ZZ$) backgrounds are
modeled with \mcatnlo interfaced to the \herwig and \Jimmy programs.
The single top background samples are produced with the AcerMC generator~\cite{AcerMC}.
$W\gamma$ is modeled using \pythia/\madgraph~\cite{Alwall:2007st}, and hadronic multi-jet backgrounds with
the \pythia generator.

The detector response for all the generated MC events is simulated by passing them
through a detailed simulation~\cite{simulation} of the ATLAS detector based on the {\tt GEANT4} program~\cite{Geant4}. 
These simulated data samples are then reconstructed, selected, and analyzed as it is done for the data.

%
%
\section{Object reconstruction}
\label{sec:ObjectReconstruction}
\indent 
Brief descriptions of the relevant physics objects used in this analysis 
are given in this section. 
These objects are the $pp$ collision vertices, electrons, muons, 
missing transverse energy, and hadronic jets.

\indent 
The $pp$ collision vertices in each bunch crossing
are reconstructed using the inner tracking system.
To remove cosmic-ray and beam-induced backgrounds,
we require the primary vertex to have three or more tracks.


\indent 
Electron candidates are formed by matching a cluster of energy in the electromagnetic
calorimeter to a charged track found in the inner tracking system. 
In this analysis, we use electrons with $|\eta| < 1.37$ or 
$1.52 < |\eta| < 2.47$, which avoids the transition region between the barrel
and the end-cap electromagnetic calorimeters.

We use the ATLAS ``tight'' electron identification
criteria~\cite{atlas-winclusive}.
The \et of the leading electron in the $ee$ channel and the
\et of the electron in the $e\mu$ channel are required to exceed 25~GeV.
The sub-leading \et in the $ee$ channel must be greater than 20~GeV. 
The isolation requirement is that the sum of the energy in the calorimeter cells within a cone 
$\Delta R = \sqrt{(\Delta\eta)^2 + (\Delta\phi)^2}$
of 0.3 around the electron, excluding the energy in the electron cluster and after corrections for
leakage and pileup effects, must be less than 4~GeV. 
In addition, the longitudinal impact parameter $|z_0|$ of the track relative to the
primary vertex must be smaller than 10 mm, 
and the transverse impact parameter significance $d_0/\sigma_{d_0}$ must
be less than 10, where $d_0$ and $\sigma_{d_0}$ 
are the transverse impact parameter relative to the primary vertex 
and its uncertainty, respectively.  
The electron identification efficiency is measured in data using control
samples of $Z\rightarrow e^+e^-$ and $W^\pm \rightarrow e^\pm\nu$ events. 
The overall electron selection efficiency in this analysis 
is about 78\% for the central region ($|\eta|<0.8$), 
and decreases to about 64\% in the endcap region ($2.0< |\eta| < 2.47$). 


\indent Muons are reconstructed using information from the muon spectrometer,
the inner tracking detectors, and the calorimeters.  
The absolute value of the pseudorapidity 
$|\eta|$ of the combined muon is required to be less than 2.4. 
The muon transverse momentum measured by combining information from the 
muon spectrometer and inner detector systems
must be greater than 20~GeV. 
To suppress muons originating from hadronic jets, 
the muons must be isolated such that the \pt sum of other tracks 
in a cone of $\Delta R$ = 0.2 around the muon track divided
by the muon \pt is less than 0.1. 
The muon reconstruction and identification efficiencies are measured using
$Z \rightarrow \mu^+ \mu^-$ candidates in data.
The overall muon reconstruction efficiency is found to be 
92.8\% $\pm$ 1.0\% (syst) in data with negligible statistical uncertainty.


\indent Jets are reconstructed from calorimeter clusters using the
anti-$k_{t}$ algorithm~\cite{antikt-jet,Cacciari:2005hq,Fastjet} 
with a jet resolution parameter of $\Delta R=0.4$. 
These jets are calibrated using \pt- and $\eta$-dependent 
correction factors~\cite{ATLAS-CONF-2011-032} based on MC simulation, and validated 
by extensive test beam and collision data. 
The selected jets are required to have $\pt>30$~GeV and $|\eta|<4.5$. 
The jet energy scale uncertainty depends on \pt and $\eta$ and ranges from
2\% - 8\% for $\pt>30$~GeV and $|\eta|<4.5$ due to
calibration~\cite{ATLAS-CONF-2011-032}. 


\indent The missing transverse energy is reconstructed using calorimeter energies
from clusters in the detection range $|\eta| < 4.5$ 
and muon momenta measured by the muon spectrometer and inner detector.  
The mis-measured energies of leptons or jets can produce false \met. 
To reduce the rate of backgrounds that arise from these
mis-measurements, particularly the Drell-Yan background, we use a
relative missing transverse energy (\metrel), which is defined as the following:
\begin{equation}
\label{eqn:MetRelDef}
\metrel = \left\{ 
\begin{array} {ll}
\met \times \sin\left(\Delta\phi_{\ell, j}\right) & \mbox{if} \; \Delta\phi < \pi/2 \\
\met                                            & \mbox{if} \; \Delta\phi \geq \pi/2,
\end{array}
\right.
\end{equation}
where $\Delta\phi_{\ell, j}$ is the difference in the azimuthal 
angle $\phi$ between the \met and the nearest lepton or jet.
The asymmetric requirement on \metrel helps to optimize the
WW detection significance. 
This requirement also reduces the background from 
$Z \rightarrow \tau^+ \tau^-$, where the real \met from the 
$\tau$ semileptonic decays is parallel to the momenta of the leptons.
\section{Event selection}
\label{sec:evt_sel}
\indent 
Di-lepton events are selected using the lepton identification requirements 
described in the previous section.
For each di-lepton final state ($ee$, or $\mu\mu$, or $e\mu$), 
we require each electron (or muon) to have \et (or \pt) $>$ 20~GeV, 
and we require exactly two leptons with opposite charges.
The electron in the $e\mu$ channel and the leading electron in the $ee$ channel  must have $E_T>25$ GeV.
Both leptons are required to originate from the primary vertex.
At least one of the selected leptons is required to have fired its
corresponding trigger (``trigger matching'').
The leptons must also have a minimum separation
(effectively imposed by the lepton isolation requirement) 
in $\eta-\phi$ space of $\Delta R > 0.3$ in the $ee$ channel 
and $\Delta R  > 0.2$ in the $\mu\mu$ and the $e\mu$ channels.



To reduce the Drell-Yan background and the background from hadronic multi-jets,
the invariant mass of the dilepton pair ($e^+e^-$ or $\mu^+\mu^-$) is required to exceed 15 GeV 
and is not allowed to be within $\pm$15 GeV of the $Z$-mass.
In addition, the invariant mass of the dilepton pair ($e^-\mu^+$ or $e^+\mu^-$) 
is required to exceed 10 GeV.

As shown in Fig.~\ref{fig:met-rel-ll}, 
further suppression of the remaining Drell-Yan and other backgrounds is achieved 
by requiring $\metrel > $40 GeV for the $ee$ channel, $\metrel > $45 GeV for the $\mu\mu$ channel 
and $\metrel > $ 25 GeV for the $e\mu$ channel. To optimize the signal to background ratio, different \metrel~cuts are
chosen for the different channels.
The figure shows the \metrel distribution 
prior to applying the jet-veto selection criteria for the 
$ee$, $\mu\mu$ and $e\mu$ channels. 

 \begin{figure}[h]
 \vspace*{-0.2cm}
 \begin{center}
 \includegraphics[width=0.4\textwidth]{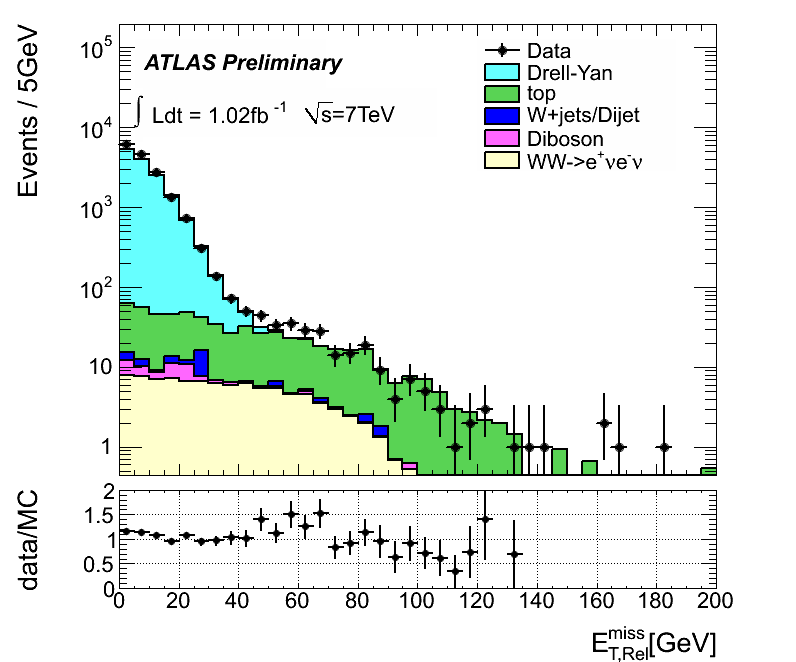}
 \includegraphics[width=0.4\textwidth]{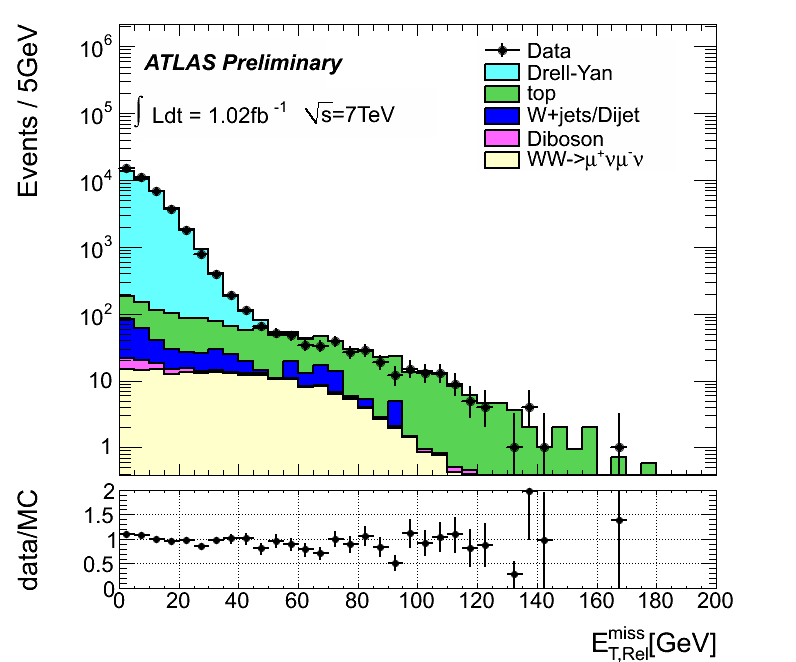}
 \includegraphics[width=0.4\textwidth]{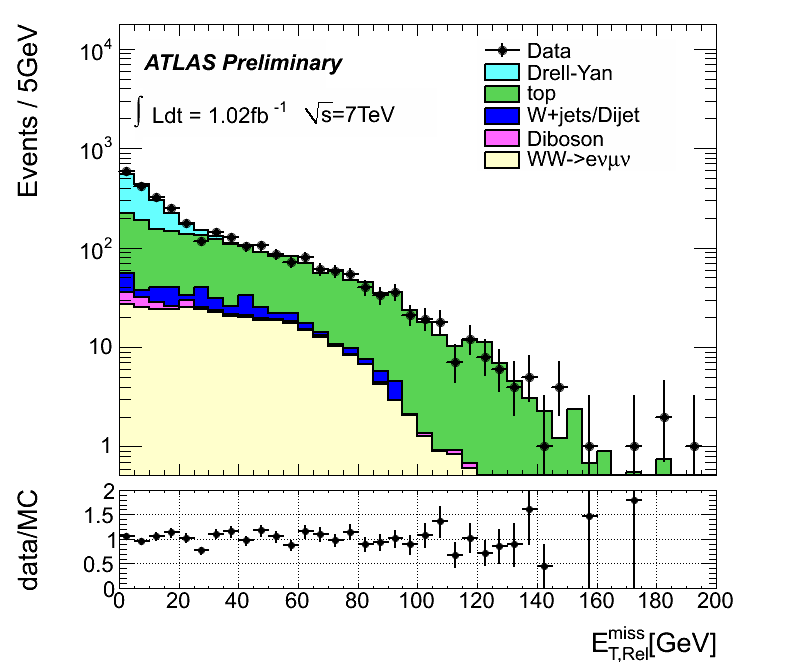}
 \includegraphics[width=0.4\textwidth]{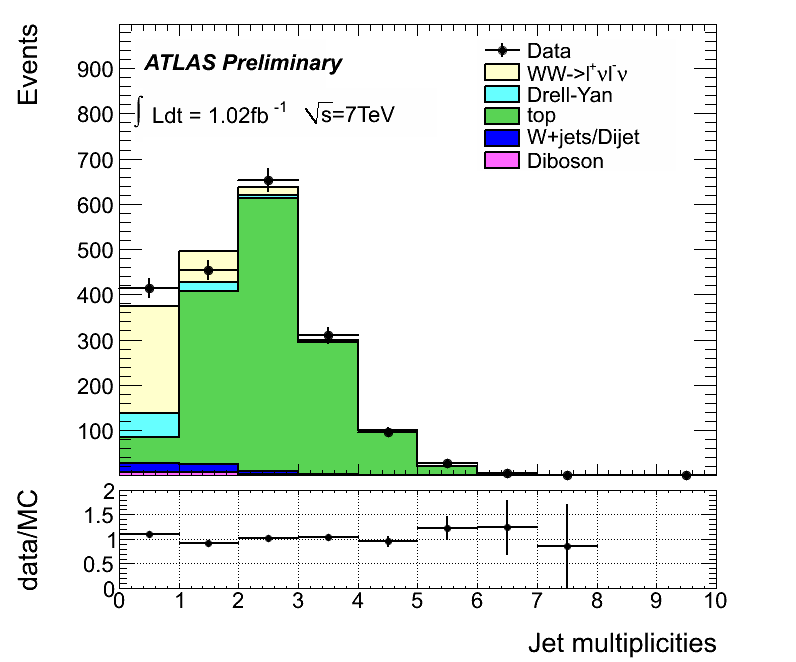}
 \caption{\small \metrel distributions for the selected $ee$, $\mu\mu$ and $e\mu$
          samples after the $Z$ mass veto cut. Bottom right plot shows the
          jet multiplicity after \metrel cut.}
 \label{fig:met-rel-ll}
 \end{center}
 \end{figure}

Suppression of the top background is accomplished by rejecting events containing 
jets with $\pt > 30$ GeV and $|\eta|<4.5$ (the jet-veto requirement).
Figure~\ref{fig:met-rel-ll} shows that the majority of the \WW signal is in 
the zero-jet bin of the jet multiplicity
distribution, while the top events populate the higher jet multiplicity bins. 

 

After all the selection criteria are applied to the di-lepton samples, 414 \WW candidate events remain: 
74 in the $ee$ channel, 97 in the $\mu\mu$ channel, and 243 in the $e\mu$ channel. 
%
%

\indent 
The signal acceptance is one of the key ingredients for determining 
the \WW cross section.
Determining this acceptance relies on detailed MC signal modeling 
and event selection efficiency corrections using control data samples.
These corrections differ by at most a few percent from
unity, indicating the inherent accuracy of the simulation.
Detailed information on the \WW event selection acceptance is provided below.
Table~\ref{ta:cutflow_mc} shows the number of MC \WW events passing
each selection cut for the three di-lepton channels. 
The MC events shown in the table are normalized to 1.02 fb$^{-1}$ using the
SM \WW production cross sections with $q\bar q'$ and $gg$ initial states. 
In order to minimize the systematic uncertainty due to initial state radiation (ISR)
and jet-energy scale,
we use control samples of $Z\rightarrow \ell^+\ell^-$ data and MC
to determine the jet-veto efficiency correction factor
to the \WW selection~\cite{Campbell:2009kg}: 
$\epsilon_{WW}^{\mathrm{data}} = \epsilon_{WW}^{\mathrm{MC}}\times f_Z,$ 
with $f_Z = \epsilon_Z^{\mathrm{data}}/\epsilon_Z^{\mathrm{MC}}=0.99\pm0.05$ (syst).
The two efficiency numbers, $\epsilon_Z^{\mathrm{data}}$ and $\epsilon_Z^{\mathrm{MC}}$,
are the fraction of the $Z$ events with
zero jets found in data and MC, respectively.
The main contributions to the given systematic uncertainty 
are due to factorisation and renormalisation scale choice
evaluated using \mcatnlo and MCFM~\cite{MCFM}.
The overall event selection acceptances for signal events are 
5.0\% for $e\nu e\nu$, 8.9\% for $\mu\nu\mu\nu$, and 12.6\% for $e\nu\mu\nu$ events.
The contributions from $W^+W^- \rightarrow \tau^\pm \nu \ell^\mp \nu \rightarrow \ell'^\pm \ell^\mp + n\nu$ 
are less than 8\% of the final selected \WW signal events
in all three di-lepton channels.
The overall systematic uncertainty in the \WW selection acceptance 
is 6.8\% for the combination of the three di-lepton channels.
This overall uncertainty includes 
the uncertainties due to lepton identification (4.2\%), the jet-veto (4.8\%), 
the \met uncertainty (2.2\%) and the uncertainties of the PDFs 
used in MC modeling (1.3\%). 
The PDF uncertainty includes the 
error matrices of the CTEQ6.6 PDF sets 
and the differences between the CTEQ6.6 and the MSTW~\cite{MSTW} PDF sets.


\begin{table}[htb] 
  \vspace{-0.2cm}
  \centering
  \begin{tabular}{lrrrrrr} \hline \hline
    Cuts & \multicolumn{2}{c}{$ee$ Channel} & \multicolumn{2}{c}{$\mu\mu$ Channel} & \multicolumn{2}{c}{$e\mu$ Channel} \\ \cline{2-7}
                              & $e\nu e\nu$  & $\tau\nu \ell\nu$& $\mu\nu \mu\nu$ & $\tau\nu \ell\nu$& $e\nu \mu\nu$ & $\tau\nu \ell\nu$ \\ \hline

    Total Events              & 552.3 & 211.4  & 552.3 & 211.4  & 1104.5 & 423.1  \\ \hline

    2 leptons (SS+OS)         & 116.6 &  11.8 & 229.0 &   25.5 & 332.7 &   35.5 \\

    2 leptons (OS)            &  115.7 &  11.6 & 229.0 & 25.5 & 331.3 & 35.3   \\

leading electron Pt $>$ 25GeV &  114.4 &  11.4 & - & - & 305.5 & 30.2  \\

trigger matching              & 114.2   & 11.4  & 231.9 & 25.8 &305.3  & 30.2   \\  \hline

$M_{\ell\ell}>$ 15 GeV, $M_{e\mu}>$ 10 GeV  &  113.5 & 11.3 & 229.7 & 25.6 & 304.5 &  30.1 \\

    Z mass veto               & 88.2  &  8.4 & 176.6 &   19.0 & - &  - \\

    \metrel cut               & 38.6  & 2.9   & 69.7  &   5.2 & 193.2 &   16.1 \\

     Jet veto (No. of Jet=0)  & 27.8  &  1.7 & 49.4  &   3.1  &  139.6 &   10.9 \\ \hline

     \WW Acceptance           & 5.0\%&  0.8\%& 8.9\%& 1.5\% & 12.6\%& 2.6\% \\ \hline

  \end{tabular}
  \caption{\WW MC event selection cut flow and overall acceptance. 
The MC \WW signal expectations are normalized to 1.02 fb$^{-1}$
integrated luminosity, using the NLO SM cross sections. $\ell$ refers to $e$, $\mu$ and $\tau$ in this table.} 
\label{ta:cutflow_mc} 
\end{table}

The systematic uncertainties on the \WW\ signal acceptance are summarized in 
Table~\ref{tab:syst_summary}, which gives the sources and associated uncertainties in \WW\ signal acceptance 
for the $ee$, $\mu\mu$ and $e\mu$ channels.

\begin{table}[htb]
  \vspace*{-0.2cm}
  \centering

  \begin{tabular}{lccc} \hline \hline
    \toprule
     Sources  & $e^+e^- \met$ & $\mu^+\mu^- \met$ & $e^\pm\mu^\mp \met$ \\ \hline
    \midrule
    Luminosity             & 3.7\% & 3.7\% & 3.7\% \\

    \midrule

    Cross-section (theory) & 5\%   &  5\%  & 5\%   \\

     \hline

    PDF                          & 1.2\%  & 1.4\%  & 1.4\% \\ 
\hline

    Trigger                      & 1.0\% & 1.0\% & 1.0\%  \\
   
    Lepton $p_T$ smearing        & 0.2\% & 0.1\% & 0.1\% \\

    Reco eff. scale factors                 & 1.4\% & 0.0\% & 0.7\%  \\

    $E_T$/$p_T$ scale correction & 0.9\% & 0.0\% & 0.4\% \\

    Particle ID eff. scale factors          & 3.3\% & 1.4\%  & 1.6\% \\

    Isolation                    & 4.0\% & 2.0\% & 3.0\% \\

    \metrel in-time contribution        & 3.5\% & 3.9\% & 1.4\% \\

    \metrel out-of-time contribution       & 0.5\%  & 0.5\%  & 0.3\% \\

    Jet-veto                     & 4.8\%  & 4.8\%  & 4.8\% \\

    Total experimental uncertainty & 8.1\% & 6.7\% & 6.2\% \\  \hline

    Overall uncertainty            &        &        &        \\       
    for WW signal estimation       & 10.3\% & 9.2\% & 8.9\% \\ \hline

    \bottomrule
  \end{tabular}

  \caption{Uncertainty sources and associated relative uncertainties for WW signal 
           for the $ee$, $e\mu $ and $\mu\mu$ channels.}
  \label{tab:syst_summary}
\end{table}
\section{Background estimation}

\indent 
The main backgrounds for the \WW signal come from Drell-Yan events, 
top ($t \bar t$ and single top), $W+$jets,  
and diboson ($WZ$, $ZZ$, $W\gamma$ and $Z\gamma$) production, as described
in this section.


\indent Drell-Yan events ($Z/\gamma^* \rightarrow \ell^+\ell^-$), 
like \WW events, produce two high \pt leptons.
Much of this background is removed by the di-lepton invariant mass cuts 
and \metrel cuts.
Given the relatively large cross section of the Drell-Yan process,
it contributes a non-negligible background to the \WW signal  
due to energy/momentum mis-measurements of the two leptons or as a result of  
hadronic activity in the rest of the event.  Extra pileup collisions
are a significant contribution to this hadronic activity.

The Drell-Yan backgrounds are determined from MC events generated with \alpgen 
and passed through the ATLAS detector simulation. 
The systematic uncertainties are determined using a data-driven method which
compares the \metrel distributions in data and MC within the $Z$ peak
region. The distributions are found to be consistent within statistical
uncertainties. As a conservative estimate of the systematic, the observed
difference is added in quadrature with the statistical uncertainty. This
systematic uncertainty is applied to the background predictions from MC.
With the assumptions that the mechanism that causes 
a discrepancy between data and MC is independent of the invariant mass of the two leptons, 
and that any discrepancy between data and MC is caused by a mis-modeling of 
the Drell-Yan sample, a systematic uncertainty of 10.4\% for the Drell-Yan background 
estimation from the MC simulation is determined. 
The estimated Drell-Yan backgrounds together with statistical and systematic 
uncertainties are shown in Table~\ref{ta:selected_data_MC}.




\indent After removing the largest background from $Z/\gamma^* \rightarrow \ell^+\ell^-$ events 
in di-lepton samples, the dominant background is from
$t\bar t $ and $Wt $, followed by the leptonic decays of two real $W$ bosons. 
This background can be effectively suppressed by requiring that there are
no jets with $\pt > 30$ GeV and $|\eta|<4.5$.
However, some top events containing only jets with
\pt less than 30 GeV can still mimic the SM \WW events. 

The top backgrounds in the final \WW selection are estimated using a data driven
method which is cross-checked with the Monte Carlo simulation.
In this method, the top background in the zero jet bin is estimated using the number of 
observed events in the $N$-jet bins (where $N\geq 2$) in data 
and the ratio of zero-jet events over the number of $N$-jet events in MC,
$N_{\mathrm{top}}^{\mathrm{zero-jet}}({\mathrm{ estimate}}) = N_{\mathrm{data}}^{\geq 2\mathrm{-jets}}\times (N_{\mathrm{MC~top}}^{\mathrm{zero-jet}}/N_{\mathrm{MC~top}}^{\geq 2\mathrm{-jets}}).$
The uncertainties on luminosity and on the top cross sections cancel out in the MC
ratio. A second method uses a top control sample selected using $b$-tagging to estimate the top contribution. 
Both data-driven methods give top background yields consistent with MC and with each other.
The final top background estimated 
is $58.6\pm2.1$(stat)$\pm 22.3$(syst) events, the systematic uncertainty 
is dominated by the jet-energy scale uncertainty (37\%).

$W$ bosons produced in association with a hadronic jet
give rise to \WW backgrounds when the jet is misidentified as a lepton.
The rate at which hadronic jets are misidentified as leptons may not be accurately 
described in the MC. The $W +$ jets background is therefore determined from data.
The $W +$ jet background is estimated by defining a control region, 
similar to the \WW signal selection, that is enriched in W+jet events.
The $W +$ jet control region is defined using an alternative lepton definition 
that is rich in hadronic jets. Jet-rich electrons are defined as isolated 
electromagnetic clusters matched to a track in the inner detector that 
fail the {\tt medium} \cite{atlas-winclusive} electron identification requirements.
Jet-rich muons are defined as ``combined'' muons that pass a looser isolation 
requirement and fail the muon selection.
Events containing one fully identified lepton and a jet passing this jet-rich lepton 
definition are selected.
These events are then required to pass the full \WW event selection, 
where the jet is treated as if it were a fully identified lepton.
The $W +$ jets background is estimated by scaling the control sample
($N_{W+\mathrm{jet}}$) by a measured fake factor.
The fake factor ($f$) is defined as the ratio of the rate at which jets satisfy 
the full lepton identification to the rate at which they satisfy the jet-enriched 
lepton selection.  The fake factor is measured in a di-jet data sample,
parametrized as a function of lepton \pt, and is found to be 0.02
(0.25) for electrons (muons) in the 30-40 GeV region.
The $W+$ jets background is determined as $N_{W+\mathrm{jet}} \times f$.
Samples with same sign di-leptons are also enhanced in $W +$ jets
events and are used as a control sample, with looser lepton
identification, to cross check the method described above.  The $W +$
jets prediction based on the fake factor method agrees well with the
observed same sign di-lepton data.
The $W+$ jets background estimation in the \WW signal region is given 
in Table~\ref{ta:selected_data_MC} for the three di-lepton channels.
The statistical uncertainty is due to the limited statistics 
in the $W+$ jet control regions.
The systematic uncertainty is due to the uncertainty on the determination of $f$.
The uncertainty on $f$ is about 30\% for both electrons and muons, and 
is determined by the measured variations of $f$ in different run periods 
and in data samples containing jets of different energies.
The assigned systematic uncertainty covers variation of the quark/gluon 
composition of the jets in the jet data sample compared to jets in the $W +$ jets sample, 
and the effects of changing instantaneous luminosity on $f$.
The total $W +$ jets contribution to the final selected $WW$ candidate events is 
estimated to be $50.5\pm4.8$(stat)$\pm14.7$(syst). 

%




\indent Other backgrounds to \WW originate from the diboson processes $WZ$, $ZZ$ and $W+\gamma$. 
The leptonic decays of $WZ$ and $ZZ$ events can mimic the \WW signal when one or more of the
charged leptons is not reconstructed and instead contributes to \met. 
The $ZZ \rightarrow \ell\ell\nu\nu $ process is suppressed by the $Z$ veto cut.
The $W\gamma$ process is a background only for the $ee$ and $e\mu$ channels, since the probability 
for a photon to be misidentified as a muon is negligible. 
These diboson background contributions are estimated using MC simulations
(\mcatnlo for $WZ$, \pythia for $ZZ$ and \pythia/\madgraph for $W+\gamma$).
The total diboson background contribution is estimated to be
6.8 $\pm$ 0.4 (stat) $\pm$ 0.8 (syst) events for 1.02 fb$^{-1}$ of integrated luminosity.
The quoted systematic uncertainty includes uncertainties
in the luminosity (3.7\%), the SM diboson cross-sections (5\%),
the jet veto efficiency (9.6\%),
the di-lepton trigger and identification efficiencies (4.5\%) 
and \met uncertainty (2.2\%).
The hadronic dijets background contribution is negligible in the \WW signal region as determined by \pythia MC
and cross-checked by data.  
\section{WW Production and Fiducial Cross Sections}

\indent The observed and expected number of events after applying all \WW selection cuts
are shown in Table~\ref{ta:selected_data_MC}.
Both statistical and systematic uncertainties are given for 
all three di-lepton channels in the table. 
\begin{table}[htb]
  \vspace*{-0.2cm}
  \centering

{
  \begin{tabular}{lcccc}
    \hline \hline
     Final State  & $e^+e^- \met$ & $\mu^+\mu^- \met$ & $e^\pm\mu^\mp \met$ & Combined \\
    \midrule
    Observed Events  & 74 & 97 & 243 & 414 \\ \hline

    \midrule

    Background estimations  &                          &                           &                          \\

     Top(data-driven)        & 9.5$\pm$0.3$\pm$3.6 & 12.3 $\pm$0.4$\pm$4.7   & 36.8$\pm$1.3$\pm$14.0 & 58.6$\pm$2.1$\pm$22.3   \\

     W+jets (data-driven)    & 5.3$\pm$0.4$\pm$1.7 & 12.4$\pm$2.9$\pm$5.2 & 32.9$\pm$3.8$\pm$9.2   & 50.5$\pm$4.8$\pm$14.7   \\

     Drell-Yan (MC/data-driven)& 18.7$\pm$1.9$\pm$1.9 & 19.2$\pm$1.7$\pm$2.1    & 16.0$\pm$2.8$\pm$1.7 & 54.0$\pm$3.7$\pm$ 4.5 \\

     Other dibosons (MC)     & 0.9$\pm$0.1$\pm$0.1  & 2.4$\pm$0.2$\pm$0.3  & 3.4$\pm$0.3$\pm$0.4 & 6.8$\pm$0.4$\pm$0.8  \\ \hline

     Total Background        & 34.4$\pm$2.0$\pm$4.4   & 46.3$\pm$3.4$\pm$7.3 & 89.1$\pm$4.9$\pm$16.8 & 169.8$\pm$6.4$\pm$27.1   \\

    \midrule
      Expected $WW$ Signal & 29.5$\pm$0.3$\pm$3.0 & 52.5$\pm$0.4$\pm$4.9 & 150.5$\pm$0.7$\pm$13.4 & 232.4$\pm$0.9$\pm$21.5   \\

     \hline \hline

     Significance ($S / \sqrt{B}$)     & 5.0            & 7.7            & 15.9 & 17.8                     \\

    \hline \hline
  \end{tabular}
}  
  \caption{\small Summary of observed events and expected signal and background contributions
           in the three di-lepton and combined channels.
           The first error is statistical, the second systematic.
The central value and statistical uncertainty for the Drell-Yan process estimation is 
MC based while the systematic uncertainties are derived from a data-driven method. 
}
  \label{ta:selected_data_MC}
\end{table}
The kinematic distributions of the final \WW candidates together with the predicted
\WW signal and estimated background are shown in Figures~\ref{fig:ww-1}.
The distributions of transverse mass 
$\mt = \sqrt{(\et^{\ell 1} + \et^{\ell 2}+\met)^2-(\vec{p}_{\mathrm{T}}^{\ell 1} + \vec{p}_{\mathrm{T}}^{\ell 2} + \vec{E}^{\mathrm{miss}}_{\mathrm{T}})^2}$ 
and transverse momentum $\pt (\ell \ell \met)$ of the di-lepton plus \met system,
where $\et^\ell $ and $\vec{p}_{\mathrm{T}}^\ell $ denote transverse energies and momenta for leptons,
and \met and $\vec{E}^{\mathrm{miss}}_{\mathrm{T}}$ denote missing transverse energy and momentum in the event.
The MC distributions in these plots are normalized to the integrated luminosity 
of 1.02 \ifb and the SM cross sections with event selection acceptance. 
The data agree reasonably well with the estimated signal and background. 

 \begin{figure}[h]
 \vspace*{-0.2cm}
 \begin{center}
 \includegraphics[width=0.3\textwidth]{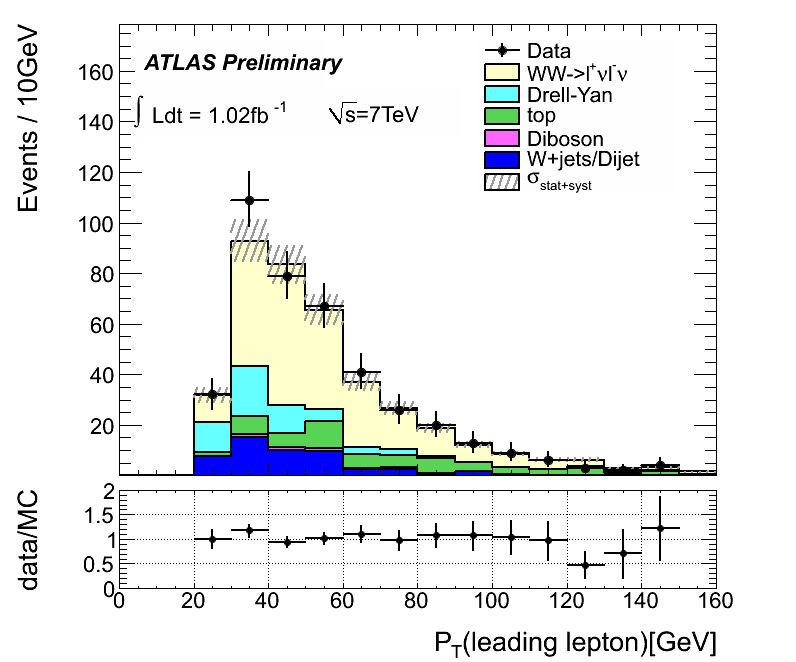}\hfill
 \includegraphics[width=0.3\textwidth]{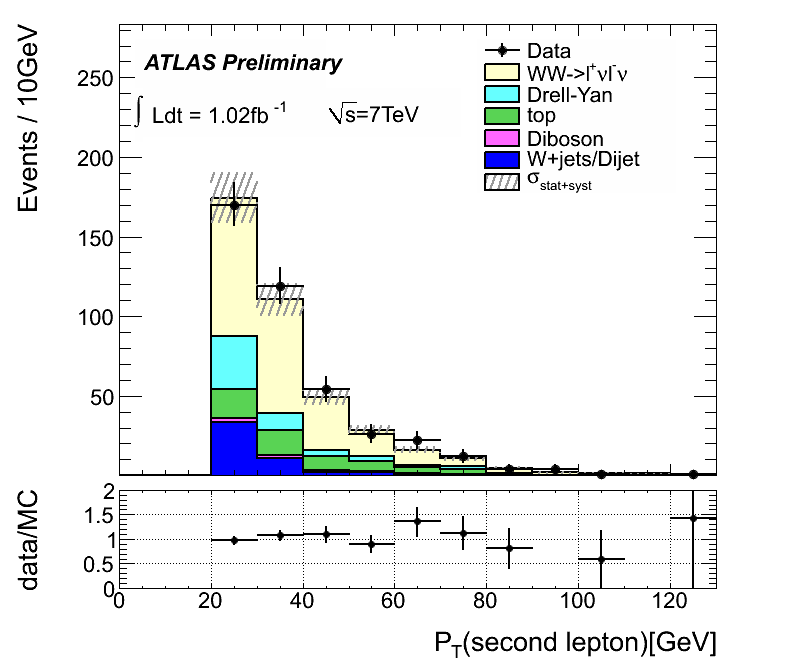}\hfill
 \includegraphics[width=0.3\textwidth]{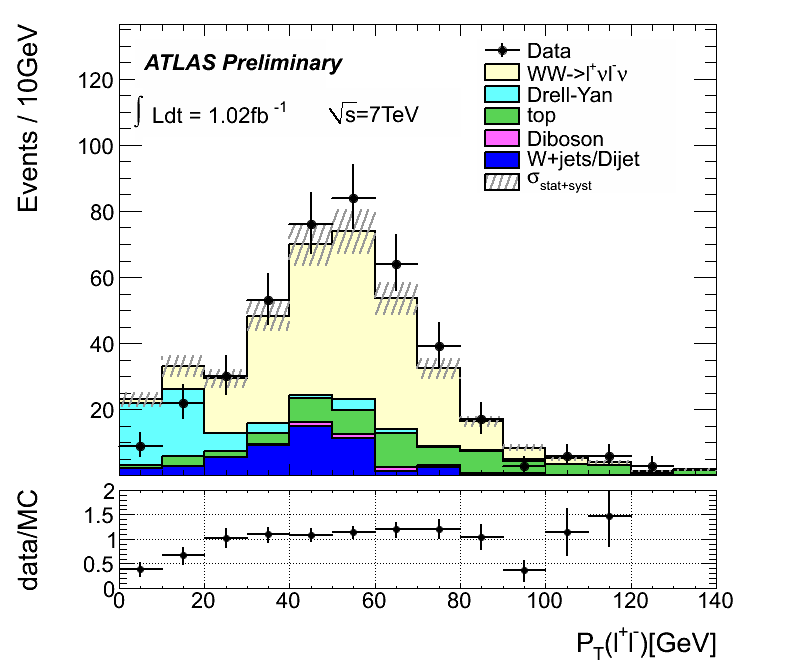}
 \includegraphics[width=0.3\textwidth]{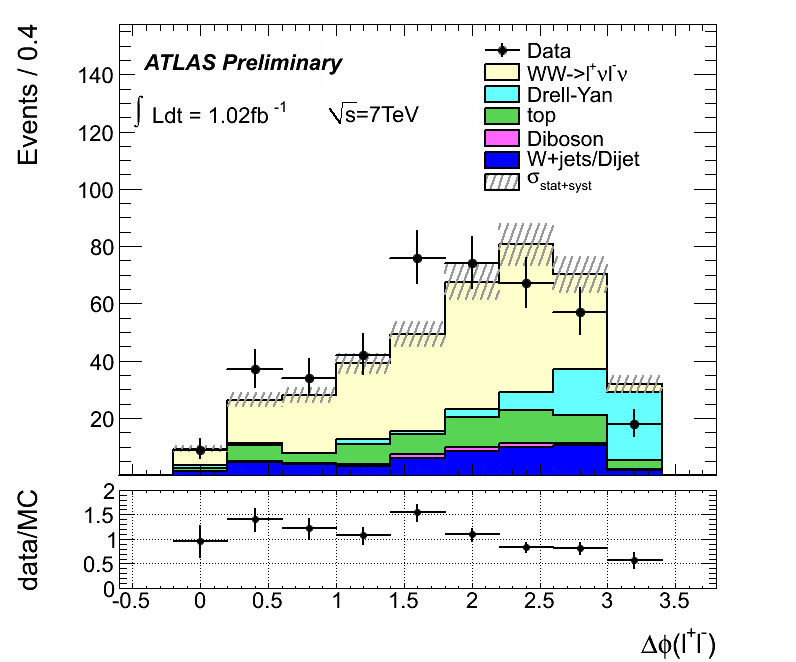}\hfill
 \includegraphics[width=0.3\textwidth]{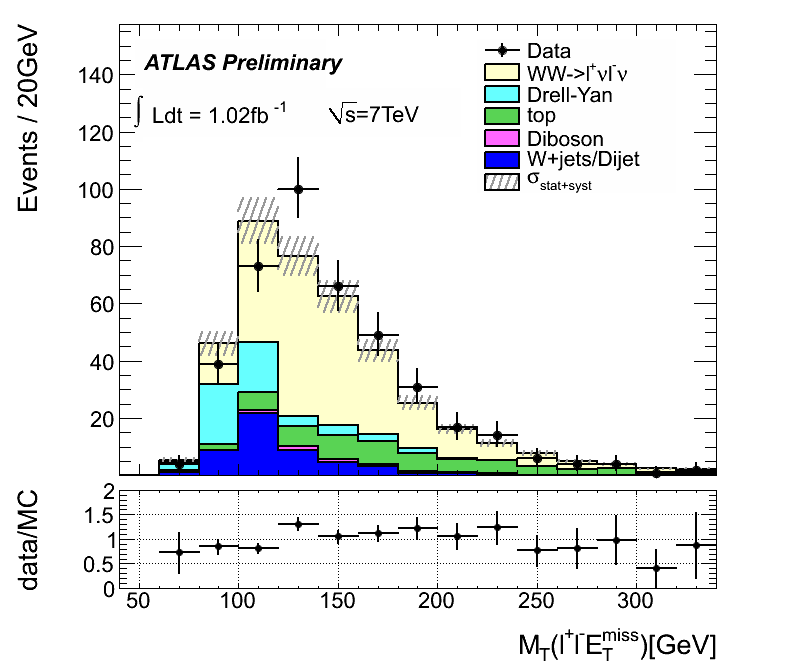}\hfill
 \includegraphics[width=0.3\textwidth]{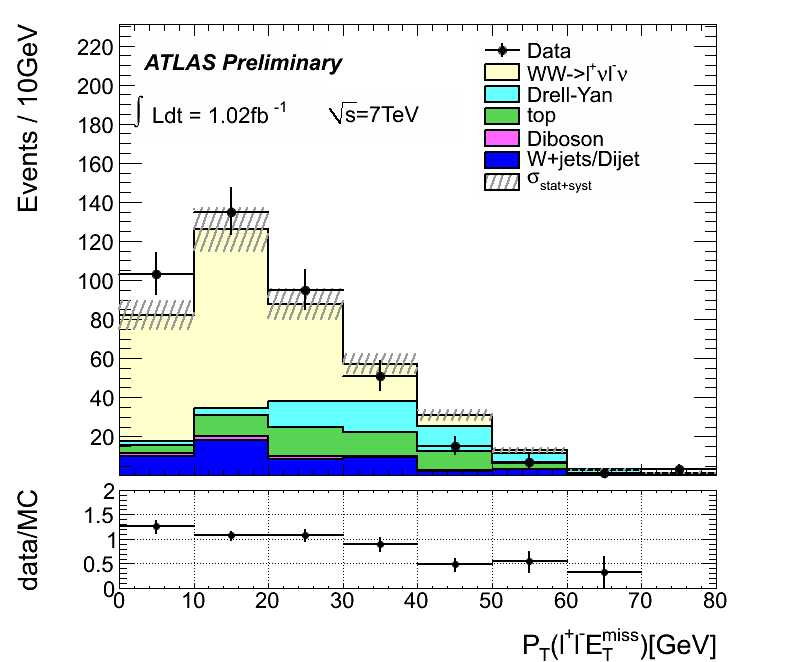}
 \caption{\small Distributions of leading (top left), sub-leading (top middle) lepton \pt 
and di-lepton system \pt (top right) for \WW candidates. 
Distributions of $\Delta\phi (\ell^+\ell^-)$ (bottom left), WW transverse mass (bottom middle) 
and \pt of WW (bottom right).
   The points are data 
   and stacked histograms are from MC predictions except the $W+$jets background (from data-driven methods).
   The estimated uncertainties are shown as the hatched bands (stat $\oplus$ syst).
   }
 \label{fig:ww-1}
 \end{center}
 \end{figure}


The total and fiducial \WW production cross sections are
determined from the three dilepton channels ($WW\rightarrow e\nu e\nu$,
$\mu\nu \mu\nu$ and $e\nu \mu\nu$) by maximizing the log-likelihood functions shown in 
Equation~\ref{eq:wwxsec} and ~\ref{eq:wwxsecfid}, respectively:
\begin{equation}
L(\sigma^{tot}_{WW}) = \ln \prod^3_{i=1} \frac{e^{-(N^i_s+N^i_b)} \times (N^i_s+N^i_b)^{N^i_{obs}}}{N^i_{obs}!},~
~ N^i_s = \sigma^{tot}_{WW} \times Br^i \times {\cal L} \times \epsilon^i_{WW}
\label{eq:wwxsec}
\end{equation}
\begin{equation}
L(\sigma^{i,fid}_{WW}) = \ln \frac{e^{-(N^i_s+N^i_b)} \times (N^i_s+N^i_b)^{N^i_{obs}}}{N^i_{obs}!},~
N^i_s = \sigma^i_{WW \rightarrow \ell\nu \ell\nu} \times {\cal L} \times C^i_{WW}
\label{eq:wwxsecfid}
\end{equation}
\noindent where $i=1,2,3$ runs over the three dilepton channels.
$N^i_s$, $N^i_b$ and $N^i_{obs}$ represent the SM expected signal, estimated
background events and observed data events for the $i$-th dilepton channel
while $Br^i$ is the branching ratio (including the contribution via $\tau$ decays) 
and ${\cal L}$ is the total integrated luminosity.
The overall correction factor required to evaluate the fiducial cross section is defined as $C_{WW}$, 
and takes into account the detector resolution, efficiency and background corrections relating 
the reconstructed event selection to the true particle-level phase space region defining the cross section. 
The overall detector acceptance described in the total cross section measurement section relates 
the fiducial phase space acceptance $A_{WW}$ and the correction factor $C_{WW}$:
$\epsilon_{WW} ~= ~A_{WW} ~\times ~C_{WW}.$
Table~\ref{ta:cutflow_mc} gives the overall efficiencies for prompt $ee$, $\mu\mu$ and $e\mu$
decays from \WW. It also gives the detection efficiencies involving the electron or muon
decay from $\tau$s. 

The mean values of the total \WW cross sections obtained from the log-likelihood maximization are listed 
in Table~\ref{ta:wwxsec}, and they are consistent with the SM NLO prediction 
for \WW production of 46$\pm$3~pb.
The statistical uncertainty is determined from the likelihood fit, and 
the systematic uncertainty of our  measurement is 13.4\%, which includes the signal acceptance uncertainty 
($\Delta\epsilon_{WW}/\epsilon_{WW}$) of 6.8\%
and the uncertainty on the background estimation  ($\Delta N_{bkg}/(N_{obs} - N_{bkg})$) of 11.5\%.
The systematic error is calculated using error propagation:
$(\Delta\sigma/\sigma)_{syst} = \sqrt{((\Delta \epsilon_{WW} /\epsilon_{WW})^2+(\Delta N_{bkg}/(N_{obs}-N_{bkg}))^2}.$
The systematic uncertainties on the cross section measurements from each dilepton channel are listed
in Table~\ref{ta:wwxsec}. 
The luminosity contributes 3.7\% systematic uncertainty to the cross-section measurements and it 
is listed separately in the table. By combining all three dilepton channels, the statistical error
(8.3\%) is smaller than the systematic uncertainty (13.4\%) in this measurement.
\begin{table}[!h]
  \vspace*{-0.2cm}
  \begin{center} 
  \begin{tabular}{l|cccc} 
  \hline \hline
    Channels         & Total cross-section (pb) & $\Delta\sigma_{stat}$(pb) &$\Delta\sigma_{syst}$(pb) & $\Delta\sigma_{lumi}$(pb) \\ \hline

    $e\nu e\nu$      & 62.1          & $\pm$ 13.5            & $\pm$ 9.1         & $\pm$ 2.3 \\ 

    $\mu\nu \mu\nu$  & 44.7          & $\pm$ 8.7             & $\pm$ 7.7         & $\pm$ 1.7 \\ 

    $e\nu \mu\nu$    & 47.3          & $\pm$  4.8            & $\pm$ 6.2         & $\pm$ 1.8 \\

    Combined         & 48.2          & $\pm$  4.0            & $\pm$ 6.4         & $\pm$ 1.8 \\ 

   \hline\hline   
  \end{tabular}
  \end{center}
  \caption{The measured total \WW production cross sections in three dilepton channels.}
  \label{ta:wwxsec} 
\end{table}






In order to minimize the extrapolation from the measured to the theoretical
cross section, the \WW fiducial phase space is defined using WW selection cuts on
MC truth information.
$A_{WW}$ denotes the acceptance for the \WW decays under
consideration (including $\tau\to e/\mu+\nu\nu$ decays), defined as the fraction of decays satisfying the
geometrical and kinematical constraints at the generator level
(fiducial acceptance). 
It is defined here after the decay leptons
emit photons via QED final state radiation; photons within a $\Delta
R<0.1$ cone are added back to the decay leptons (``dressed'' leptons).
$C_{WW}$ denotes the ratios between the total number of generated
events which pass the final selection requirements after
reconstruction and the total number of generated events within the
fiducial acceptance. This corrections factor includes the efficiencies
for triggering, reconstructing, and identifying the \WW decays falling
within the acceptance.
The systematic uncertainty of $A_{WW}$ is about 6.9\%. It includes the
parton distribution function uncertainty (1.2\% for $ee$, 1.4\% for
$\mu\mu$ and $e\mu$), the renormalization and factorization scales
uncertainty (5.3\% for $ee$, 4.4\% for $e\mu$ and 1.5\% for $\mu\mu$)
and the uncertainty due to the parton shower/fragmentation modeling 
difference between \mcatnlo and MCFM NLO MC generators (4.8\%).
The MC scale uncertainty was evaluated using MC samples where the
normalization and the factorization scales were changed to one half
and to twice their nominal value.

The overall \WW event selection efficiency can be presented as
$A_{WW}\times C_{WW}$. Using the overall efficiencies for different
channels shown in Table~\ref{ta:cutflow_mc}, we can determine the
$C_{WW}$ once the $A_{WW}$ are determined. The systematic uncertainties
from various sources for $C_{WW}$ are listed in
Table~\ref{tab:syst_summary}.
The systematic uncertainty associated
with the jet-veto cut, 4.8\%, is replaced by a 4.5\% JES uncertainty term. 
The systematic uncertainty on $C_{WW}$ related to changes in the
normalization and factorization scales is 1.9\% in the $ee$ channel,
2.9\% in the $e\mu$ channel, and 5.1\% in the $\mu\mu$ channel.
A total cross section measurement requires the evaluation of the systematic on
$A_{WW}\times C_{WW}$, for which the systematics on $A_{WW}$
and $C_{WW}$ partially overlap.
Table~\ref{ta:fideff} gives the fiducial phase space acceptance
$A_{WW}$ and the correction factor $C_{WW}$ for all three dilepton
channels. 
\begin{table}[!h]
  \vspace*{-0.2cm}
  \begin{center} 
  \begin{tabular}{lccc} 
  \hline \hline
    Channels        & $A_{WW}\times C_{WW} $    & $A_{WW} $                 &  $C_{WW}$                 \\  \hline
		    						
    $e\nu e\nu$     & $0.039\pm0.001\pm0.004$   & $0.090\pm0.001\pm0.007$    &  $0.432\pm0.006\pm0.035$  \\ 
		    						
    $\mu\nu \mu\nu$ & $0.069\pm0.001\pm0.006$   & $0.086\pm0.001\pm0.005$    &  $0.802\pm0.006\pm0.066$  \\ 
		    						
    $e\nu \mu\nu$   & $0.100\pm0.001\pm0.008$   & $0.167\pm0.001\pm0.011$    &  $0.596\pm0.005\pm0.040$  \\ \hline
		    						
  \hline
  \end{tabular}
  \end{center}
  \caption{The WW overall acceptance $A_{WW}\times C_{WW} $, fiducial
    phase space acceptance $A_{WW}$ and correction factor $C_{WW}$.  The
    first errors mean statistical errors and the second errors represent
    systematic errors.}
  \label{ta:fideff} 
\end{table}
\begin{table}[!h]
  \vspace*{-0.2cm}
  \begin{center} 
  \begin{tabular}{l|ccccc} 
  \hline \hline
    Channels  & expected $\sigma^{fid}$ (fb)& 
                measured $\sigma^{fid}$(fb) & $\Delta\sigma_{stat}$(fb) &$\Delta\sigma_{syst}$(fb) & $\Delta\sigma_{lumi}$(fb) \\ \hline

    $e\nu e\nu$      & 66.8  &  90.1         & $\pm$   18.9           & $\pm$ 11.3         & $\pm$ 3.3 \\ 

    $\mu\nu \mu\nu$  & 63.8  &  62.0         & $\pm$   12.1          & $\pm$ 10.7         & $\pm$ 2.3 \\ 

    $e\nu \mu\nu$    & 245.1 &  252.0        & $\pm$   24.6          & $\pm$ 29.4         & $\pm$ 9.3 \\
\hline\hline
  \end{tabular}
  \end{center}
  \caption{The predicted and measured fiducial \WW production cross sections.}
  \label{ta:fidwwxsec} 
\end{table}
\vspace*{-0.5cm}
\section { Summary }

The \WW production cross section in $pp$ collisions at $\sqrt s = 7~\TeV~$ is 
measured using three \WW\ leptonic decay channels and 1.02 \ifb of data 
collected by the ATLAS detector during 2011. 
A total of 414 candidates are selected with an estimated background of 
170$\pm$28 events. The measured cross section is 
$48.2 \pm 4.0 {\rm(stat)} \pm 6.4 {\rm(syst)} \pm 1.8 {\rm(lumi)}$ pb, 
consistent with the SM NLO prediction of $46 \pm 3$~pb.

\begin{acknowledgments}
The author would like to thank CERN for the very successful operation of the LHC
and the ATLAS Collaboration for excellent work on the Monte
Carlo simulation and the software package for physics analysis. 
The author is supported by the Department of Energy 
(DE-FG02-95ER40899) of the United States.	
\end{acknowledgments}

\begin{thebibliography}{99}   







\bibitem{atlas-ww2010} 
{ATLAS} Collaboration, 
\href{http://prl.aps.org/abstract/PRL/v107/i4/e041802} {Phys. Rev. Lett. 107, 041802 (2011)}.

\bibitem{detectorpaper}
{ATLAS} Collaboration, 
\href{http://dx.doi.org/10.1088/1748-0221/3/08/S08003}{JINST {3} (2008)
  S08003}.

\bibitem{ATLAS-CONF-2011-011}
{ATLAS Collaboration}, 
  {ATLAS-CONF-2011-011}.

\bibitem{MCatNLO}
S.~Frixione and B.~R. Webber, 
JHEP {06} (2002)  029,
\href{http://arxiv.org/abs/hep-ph/0204244}{{hep-ph/0204244}}.

\bibitem{CTEQ66}
P.~M. Nadolsky et al., 
Phys. Rev. {D78} (2008)  013004,
\href{http://arxiv.org/abs/0802.0007}{{arXiv:0802.0007 [hep-ph]}}.

\bibitem{herwig}
G.~Corcella et al., 
  \href{http://dx.doi.org/10.1088/1126-6708/2001/01/010}{JHEP {0101} (2001)
   010},
\href{http://arxiv.org/abs/hep-ph/0011363}{{arXiv:hep-ph/0011363}}.

\bibitem{Jimmy}
J.~M. Butterworth et al., 
\href{http://dx.doi.org/10.1007/s002880050286}{Z. Phys. {C72}
  (1996)  637--646}.

\bibitem{Binoth:2006mf}
T.~Binoth, M.~Ciccolini, N.~Kauer, and M.~Kramer, 
JHEP {12} (2006)  046,
\href{http://arxiv.org/abs/hep-ph/0611170}{{arXiv:hep-ph/0611170}}.

\bibitem{pythia}
T.~Sjostrand et al., 
\href{http://dx.doi.org/10.1016/S0010-4655(00)00236-8}{Comput.
  Phys. Commun. {135} (2001)  238--259},
\href{http://arxiv.org/abs/hep-ph/0010017}{{arXiv:hep-ph/0010017}}.

\bibitem{AcerMC}
B.~P. Kersevan and E.~Richter-Was, 
\href{http://arxiv.org/abs/hep-ph/0405247}{{arXiv:hep-ph/0405247}}.

\bibitem{Alwall:2007st}
J.~Alwall et al., 
JHEP {09} (2007)  028,
\href{http://arxiv.org/abs/0706.2334}{{arXiv:0706.2334 [hep-ph]}}.

\bibitem{simulation}
{ATLAS} Collaboration, 
  \href{http://dx.doi.org/10.1140/epjc/s10052-010-1429-9}{Eur.Phys.J. {C70}
  (2010)  823--874}, \href{http://arxiv.org/abs/1005.4568}{{arXiv:1005.4568
  [physics.ins-det]}}.

\bibitem{Geant4}
S.~Agostinelli et al., {\em {GEANT4: A Simulation toolkit}\/},
  \href{http://dx.doi.org/doi:10.1016/S0168-9002(03)01368-8}{Nucl.Instrum.Meth.
  {A506} (2003)  250--303}.

\bibitem{atlas-winclusive}
{ATLAS} Collaboration, 
  \href{http://dx.doi.org/10.1007/JHEP12(2010)060}{JHEP {12} (2010)  060},
\href{http://arxiv.org/abs/1010.2130}{{arXiv:1010.2130 [hep-ex]}}.

\bibitem{antikt-jet}
M.~Cacciari, G.~P. Salam, and G.~Soyez, 
\href{http://dx.doi.org/10.1088/1126-6708/2008/04/063}{JHEP
  {04} (2008)  063},
\href{http://arxiv.org/abs/0802.1189}{{arXiv:0802.1189 [hep-ph]}}.

\bibitem{Cacciari:2005hq}
M.~Cacciari and G.~P. Salam, 
  \href{http://dx.doi.org/10.1016/j.physletb.2006.08.037}{Phys. Lett. {B641} (2006)  57--61},
\href{http://arxiv.org/abs/hep-ph/0512210}{{arXiv:hep-ph/0512210}}.

\bibitem{Fastjet}
 M.~Cacciari, G.~P.~Salam, G.~Soyez, http://fastjet.fr/.

\bibitem{ATLAS-CONF-2011-032}
{ATLAS Collaboration}, 
  {ATLAS-CONF-2011-032}.

\bibitem{Campbell:2009kg}
J.~M. Campbell et al., 
  \href{http://dx.doi.org/10.1103/PhysRevD.80.054023}{Phys. Rev. {D80}
  (2009)  054023},
\href{http://arxiv.org/abs/0906.2500}{{arXiv:0906.2500 [hep-ph]}}.

\bibitem{MCFM}
J.~M. Campbell and R.~K. Ellis, 
\href{http://dx.doi.org/10.1103/PhysRevD.62.114012}{Phys.
  Rev. {D62} (2000)  114012},
\href{http://arxiv.org/abs/hep-ph/0006304}{{arXiv:hep-ph/0006304}}.

\bibitem{MSTW}
A.~D. Martin et.al., 
  \href{http://dx.doi.org/10.1140/epjc/s10052-009-1072-5}{Eur. Phys. J. {C63} (2009)  189--285},
\href{http://arxiv.org/abs/0901.0002}{{arXiv:0901.0002 [hep-ph]}}.


\end{thebibliography}

\end{document}